\pdfoutput=1

\documentclass[11pt]{article}

\PassOptionsToPackage{dvipsnames}{xcolor}
\usepackage{EMNLP2023}

\usepackage{times}
\usepackage[T1]{fontenc}
\usepackage[utf8]{inputenc}
\usepackage{inconsolata}

\usepackage{amsmath,amsfonts,bm}

\def\eqref#1{equation~\ref{#1}}

\def\1{\bm{1}}

\def\vm{{\bm{m}}}

\def\vx{{\bm{x}}}
\def\vy{{\bm{y}}}

\DeclareMathAlphabet{\mathsfit}{\encodingdefault}{\sfdefault}{m}{sl}
\SetMathAlphabet{\mathsfit}{bold}{\encodingdefault}{\sfdefault}{bx}{n}

\usepackage{hyperref}
\usepackage{url}
\usepackage{amsmath,amsfonts,bm}
\usepackage{tikz}
\usepackage{pgfplots}
\usepackage{paralist}
\usepackage{enumitem}
\usepackage{booktabs}
\usepackage{xspace}
\usepackage{cleveref}
\usepackage{caption}
\usepackage{multirow}
\usepackage{subcaption}

\usepackage[frozencache,cachedir=minted-cache]{minted}
\setminted{stripnl=false}

\usepackage{scalerel,xparse}

\newcommand{\cbs}{CodeBERTScore\xspace}
\newcommand{\nlcode}{NL$\rightarrow$Code\xspace}
\definecolor{myblue}{HTML}{6C8EBF}
\definecolor{mygreen}{HTML}{82B366}
\definecolor{myred}{HTML}{b85450}
\newcommand{\norm}[1]{\left\lVert#1\right\rVert}

\newcommand{\downloaded}{1,000,000\xspace}

\newcommand{\eg}{\textit{e.g.,}\xspace}

\newcommand{\fbseries}{\unskip\setBold\aftergroup\unsetBold\aftergroup\ignorespaces}

\newcommand{\he}{HumanEval\xspace}
\newcommand{\conala}{CoNaLa\xspace}
\newcommand{\share}{ShareCode\xspace}
\newcommand{\davinci}{\texttt{code-davinci-002}\xspace}
\newcommand{\kt}{Kendall-Tau\xspace}
\newcommand{\ps}{Pearson\xspace}
\newcommand{\spe}{Spearman\xspace}
\newcommand{\ktt}{\ensuremath{\tau}\xspace}
\newcommand{\pss}{\ensuremath{r_p}\xspace}
\newcommand{\spp}{\ensuremath{r_s}\xspace}

\newcommand{\fref}{\ensuremath{f^*}}

\title{\cbs: Evaluating Code Generation \\ with Pretrained Models of Code}

\author{
    Shuyan Zhou\thanks{~~Equal contribution} \quad
    Uri Alon$^*$\thanks{\;\;Now at Google DeepMind} \quad
    Sumit Agarwal \quad
    Graham Neubig \\
    Language Technologies Institute, Carnegie Mellon University\quad\quad \\
  {\tt \{shuyanzh,ualon,sumita,gneubig\}@cs.cmu.edu} \\
}

\begin{document}
\maketitle

\begin{abstract}
    Since the rise of neural natural-language-to-code models (NL$\rightarrow$Code) that can generate long expressions and statements rather than a single next-token, one of the major problems has been reliably evaluating their generated output. 
    In this paper, we propose \cbs: an evaluation metric for code generation, which builds on BERTScore \citep{zhangbertscore}.
    Instead of encoding only the generated tokens as in BERTScore, \cbs also encodes the natural language input preceding the generated code, thus modeling the consistency between the generated code and its given natural language context as well.    
    We perform an extensive evaluation of \cbs across four programming languages. We find that \cbs achieves a higher correlation with human preference and with functional correctness than all existing metrics. That is, generated code that receives a higher score by \cbs is more likely to be preferred by humans, as well as to function correctly when executed. 
    We release five language-specific pretrained models to use with our publicly available code.
    Our language-specific models have been downloaded more than \textbf{\downloaded} times from the Huggingface Hub.\footnote{The code and data are available at \url{https://github.com/neulab/code-bert-score}}
\end{abstract}

\section{Introduction}
Natural-language-to-code generation (\nlcode) 
has seen sharply growing popularity recently due to the emergence of large language models (LLMs) trained on vast amounts of natural language and code 
\citep{chen2021evaluating,fried2022incoder,zhou2022doccoder,austin2021program,allal2023santacoder}.
LLMs have reached such a high \nlcode accuracy that they are now useful for the broad programming audience and actually save developers' time when implemented in tools such as GitHub's Copilot. 
This sharp rise in LLMs' usability was achieved thanks to their ability to accurately generate \emph{long} completions, which span multiple tokens and even lines, rather than only a single next-token as in early models \citep{allamanis2013mining,movshovitz2013natural}.
Nevertheless, evaluating and comparing different models has remained a challenging problem \citep{xu2022systematic} that requires an accurate and reliable evaluation metric for the quality of the models' generated outputs, and existing metrics are sub-optimal.

\begin{figure*}[t]
  \newcommand{\lsize}{0.44\textwidth}
  \newcommand{\msize}{0.43\textwidth}
  \newcommand{\rsize}{0.55\textwidth}  
  \centering
\begin{minipage}{\lsize}
\centering
\textbf{Reference:} 
\end{minipage}\\
\begin{minipage}{\lsize}
\begin{minted}[fontsize=\footnotesize,stripnl=false,framesep=1pt,frame=single,breaksymbolleft=\;,escapeinside=||]{java}
int f(Object target) {
  int i = 0;
  for (Object elem: this.elements) {
    if (elem.equals(target)) {
      return i;
    }
    i++;
  }
  return -1;
}
\end{minted}
\end{minipage} \\
\begin{minipage}{0.8\textwidth}
  \centering
  \begin{subfigure}[b]{1\linewidth}
  \caption{The ground truth reference -- find \emph{the index} of \texttt{target} in \texttt{this.elements}.}
  \label{subfig:reference}
  \end{subfigure}
  \end{minipage} \\
    \vspace{2mm}
\begin{minipage}{\msize}
  \centering
  \textbf{Non-equivalent candidate:} 
  \end{minipage} 
  \hfill
  \begin{minipage}{\rsize}
  \centering
  \textbf{Equivalent candidate:} 
  \end{minipage} \\
\begin{minipage}{\msize}
\begin{minted}[fontsize=\footnotesize,stripnl=false,framesep=1pt,frame=single,breaksymbolleft=\;,escapeinside=||]{java}
boolean f(Object target) {
  for (Object elem: this.elements) {
    if (elem.equals(target)) {
      return true;
    }
  }
  
  return false;
}


\end{minted}
\end{minipage}
\hfill
\begin{minipage}{\rsize}
\begin{minted}[fontsize=\footnotesize,framesep=1pt,stripnl=false,frame=single,breaksymbolleft=\;,escapeinside=||]{java}
int f(Object target) {
  for (int i=0; i<this.elements.size(); i++) {
    Object elem = this.elements.get(i);
    if (elem.equals(target)) {
      return i;
    }
  }
  return -1;
}

\end{minted}
\end{minipage}

\begin{minipage}{\msize}
\centering
\begin{subfigure}[b]{1\linewidth}
\caption{\textbf{\textcolor{red}{Preferred by BLEU \& CrystalBLEU}}  -- find \emph{whether or not} \texttt{target} is in \texttt{this.elements}.}
\label{subfig:noneq}
\end{subfigure}
\end{minipage}
\hfill
\begin{minipage}{\rsize}
\centering
\begin{subfigure}[b]{1\linewidth}
\centering
\caption{\textbf{\textcolor{ForestGreen}{Preferred by \cbs}}  -- find \emph{the index} of \texttt{target} in \texttt{this.elements}.}
\label{subfig:eq}
\end{subfigure}
\end{minipage}
\caption{An intuitive example for the usefulness of \cbs in measuring generated code: \Cref{subfig:reference} shows a reference code snippet in Java. 
\Cref{subfig:noneq} and \Cref{subfig:eq} show two generated predictions.
Among these two candidates and given the reference, both BLEU and CrystalBLEU prefer (score higher) the snippet in \Cref{subfig:noneq}, which \emph{is not} functionally equivalent to the reference, while our proposed \cbs prefers the code in \Cref{subfig:eq}, which \emph{is} functionally equivalent to the code in \Cref{subfig:reference}. 
}
\label{fig:intro_example}
\end{figure*}

\paragraph{Existing evaluation approaches}
The most common evaluation metrics are token-matching methods such as BLEU \citep{papineni2002bleu}, adopted from natural language processing. These metrics are based on counting overlapping n-grams in the generated code and the reference code.
CrystalBLEU \citep{eghbali2022crystalbleu} extends BLEU by ignoring the  500 most occurring n-grams, arguing that they are trivially shared between the prediction and the reference. Nonetheless, both BLEU and CrystalBLEU rely on the lexical \emph{exact match} of tokens, which does not account for diversity in implementation, variable names, and code conventions. \Cref{fig:intro_example} shows an example: given the reference code in \Cref{subfig:reference}, both BLEU and CrystalBLEU prefer (rank higher) \emph{the non-equivalent} code in \Cref{subfig:noneq} over the functionally equivalent code in \Cref{subfig:eq}.

CodeBLEU \citep{ren2020codebleu} attempts to lower the requirement for a lexical exact match, by relying on data-flow and Abstract Syntax Tree (AST) matching as well; nevertheless,  valid generations may have different ASTs and data flow from the reference code, which may lead to low CodeBLEU score even when the prediction is correct. Further, \emph{partial} predictions may be useful for a programmer,
but accepting them may lead to partial code that does not parse, and thus cannot be fully evaluated by CodeBLEU 
(for example, predicting the first line of a \texttt{for} loop, without the loop's body).

\emph{Execution-based evaluation} attempts to address these problems by running tests on the generated code to verify its functional correctness \citep{chen2021evaluating,athiwaratkun2022multi,li2022competition,wang2022execution,lai2022ds}. This provides a direct measure of the functionality of the generated code while being agnostic to diversity in implementation and style. However, execution-based evaluation requires datasets that are provided with hand-written test cases for each example, which is costly and labor-intensive to create; thus, only few such datasets exist. 
Additionally, executing model-generated code is susceptible to security threats, 
and thus should be run in an isolated sandbox, which makes it technically cumbersome to work with iteratively.

\begin{figure*}[t]
\centering
\includegraphics[width=\linewidth]{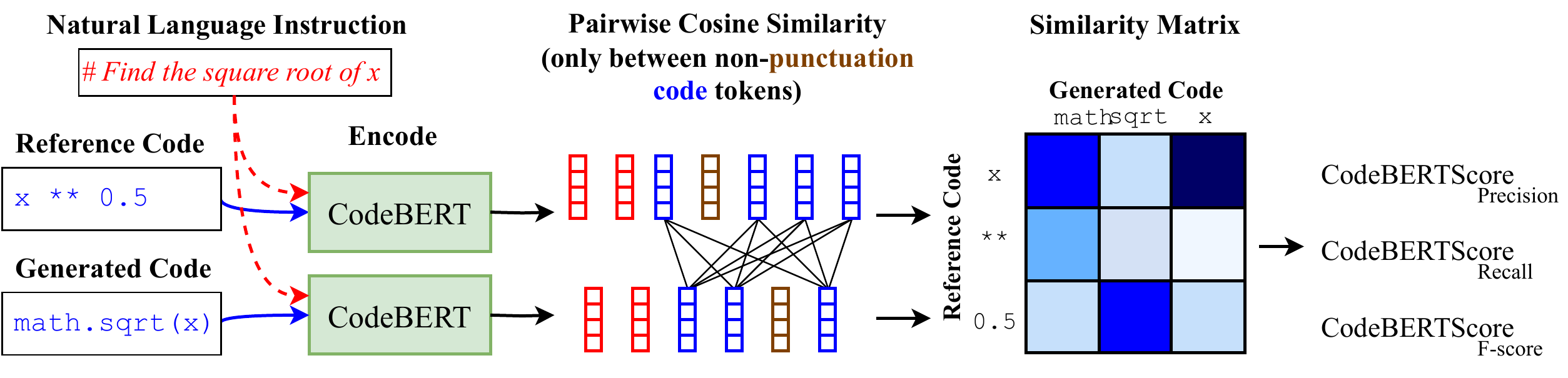}
\caption{A diagram illustrating \cbs: We use a language-specific CodeBERT model to encode each of $\langle$\emph{natural\_language}, \emph{reference\_code}$\rangle$ and $\langle$\emph{natural\_language}, \emph{generated\_code}$\rangle$. We then compute the pairwise cosine similarity between every encoded token in the reference and every encoded token in the generated code, ignoring the encoded natural language context tokens and encoded punctuation tokens; finally, we take the $max$ across the rows of the resulting matrix to compute \emph{Precision} 
and across columns to compute \emph{Recall}.
}
\label{fig:mainfigure}
\end{figure*}

\paragraph{Our approach}
In this work, we introduce \cbs, an evaluation metric for code generation, leveraging self-supervised pretrained models of code such as CodeBERT \citep{feng2020codebert}, and adopting best practices BERTScore \citep{zhangbertscore}.
First, \cbs encodes the generated code and the reference code independently with pretrained models, \emph{with} the inclusion of natural language instructions or comments.
Then, we compute the cosine similarity between the encoded representations of each token in the generated code and each token in the reference code. Finally, the best matching token vector pairs are used to compute precision and recall.
\cbs allows comparing code pairs that are lexically different while taking into account the (1) programmatic- or natural-language-context, if such provided; the (2) contextual information of each token; 
and (3) implementation diversity. 
Our approach is illustrated in \Cref{fig:mainfigure}.

\paragraph{Example}
A concrete example is shown in \Cref{fig:intro_example}: 
while BLEU and CrystalBLEU prefer
(rank higher) 
\emph{the non-equivalent} code in \Cref{subfig:noneq} given the reference code in \Cref{subfig:reference}, \cbs prefers the code in \Cref{subfig:eq}, which \emph{is} functionally equivalent to the reference (\Cref{subfig:reference}). 
We note that in this example, the variable names are identical across all three code snippets. 
When the variable names of the reference are different than the candidate's, it is \emph{even harder} for token-matching approaches such as BLEU and CrystalBLEU to compare the reference with the candidates, while \cbs can trivially match variable names according to their semantic similarity and their functional role in the code.

\paragraph{Contributions} In summary, our main contributions are:
\begin{inparaenum}[(a)]
    \item \cbs: a self-supervised metric for \nlcode evaluation, based on BERTScore,
    which leverages the benefits of pretrained models, while not requiring labeling or manually annotated data.
    \item An extensive empirical evaluation across four programming languages, showing that \cbs is more correlated with human preference \emph{and} more correlated with execution correctness than all previous approaches including BLEU, CodeBLEU, and CrystalBLEU.
    \item We pretrain and release five language-specific CodeBERT models to use with our publicly available code, for Java, Python, C, C++, and JavaScript. As of the time of this submission, our models have been downloaded from the Huggingface Hub more than \textbf{\downloaded} times.
\end{inparaenum}

\section{Evaluating Generated Code}
\subsection{Problem Formulation}
Given a context $x \in \mathcal{X}$~(\eg a natural language instruction or comment), a code generation model $\mathcal{M}: \mathcal{X} \rightarrow \mathcal{Y}$ produces a code snippet $\hat{y}\in \mathcal{Y}$ by conditioning on the intent specified by $x$. 
The quality of the generation is evaluated by comparing $\hat{y}\in \mathcal{Y}$ with the reference implementation $y^* \in \mathcal{Y}$, using a metric function $f: \mathcal{Y}\times\mathcal{Y} \rightarrow \mathbb{R}$, essentially computing $f(\hat{y}, y^*)$. 

A larger value of $f(\hat{y}, y^*)$ indicates that the generated code is more accurate with respect to the reference code,
and the way $f$ ranks different candidates is more important than the  absolute value of $f(\hat{y}, y^*)$. That is, ideally, if a prediction $\hat{y}_1$ is more functionally equivalent to $y^*$ and more preferable by human programmers over a prediction $\hat{y}_2$, we wish that a good metric would rank $\hat{y}_1$ higher than $\hat{y}_2$. That is, we seek an $f$ function such that $f(\hat{y}_1, y^*) > f(\hat{y}_2, y^*)$.

\subsection{Background: BERTScore}
BERTScore \citep{zhangbertscore} was proposed as a method for evaluating mainly machine translation outputs. 
The idea in BERTScore is to encode the candidate sentence (the prediction) and the reference sentence (the ground truth) separately, using a BERT-based model, which encodes each sequence of tokens as a sequence of vectors. Then, BERTScore computes the cosine similarity between every vector from the candidate sequence and every vector from the reference sequences. 

Given these similarity scores, BERTScore computes sentence-level precision by taking the maximum similarity score for every candidate vector and averaging, and computes recall by taking the average of the maximum similarity scores for every reference vector. 
Intuitively, a high BERTScore-recall is obtained, for example, if every vector from the reference sentence has at least one vector from the candidate sentence that is highly cosine-similar to it;  a high BERTScore-precision is obtained if every vector from the candidate sentence is highly cosine-similar to at least one vector from the reference sentence.
Ultimately, the final score is the F$_1$ score, computed as the harmonic mean of precision and recall.

\begin{figure*}[h]
\begin{align}
   & \mathrm{\cbs}_{\mathrm{P}} &&= \frac{1}{\lvert\hat{y}[\hat{\vm}]\rvert}\sum_{\hat{y}_j \in \hat{y}[\hat{\vm}]} \max_{y^*_i \ \in y^*[\vm^*]} sim\left(y^*_i,\hat{y}_j \right)  && \tag{4}\\
   & \mathrm{\cbs}_{\mathrm{R}} &&=  \frac{1}{\lvert y^*[\vm]\rvert}\sum_{y^*_i \in y^*[\vm^*]} \max_{\hat{y}_j \ \in \hat{y}[\hat{\vm}]} sim\left(y^*_i,\hat{y}_j \right) && \tag{5}\\
   & \mathrm{\cbs}_{\mathrm{F}_1} &&= \frac{2 \cdot \mathrm{\cbs}_{\mathrm{P}} \cdot \mathrm{\cbs}_{\mathrm{R}}}{\mathrm{\cbs}_{\mathrm{P}} + \mathrm{\cbs}_{\mathrm{R}}} && \tag{6}\\
   & \mathrm{\cbs}_{\mathrm{F}_3} &&= \frac{10 \cdot \mathrm{\cbs}_{\mathrm{P}} \cdot \mathrm{\cbs}_{\mathrm{R}}}{9 \cdot \mathrm{\cbs}_{\mathrm{P}} + \mathrm{\cbs}_{\mathrm{R}}} && \tag{7}
\end{align}
\caption{Main equations for \cbs}
\label{eq:main_eq}
\end{figure*}
\subsection{\cbs}
Our approach generally follows BERTScore, with the following main differences:
\begin{enumerate}[topsep=0pt,itemsep=-1ex,partopsep=1ex,parsep=1ex]
    \item We encode the context (the natural language instruction or comment) \emph{along} with each of the generated and reference code snippets, but without using the encoded context in the final similarity computation, essentially computing $f(\hat{y}, y^*, x)$ rather than $f(\hat{y}, y^*)$.
    \item Given the precision and recall, instead of computing the F$_1$ score, we also compute F$_3$  to weigh recall higher than precision, following METEOR \citep{banerjee2005meteor}.
    \item As our underlying BERT-like model, 
    we use programming language-specific models that we pretrain and release,
    rather than models that were intended for natural language only.
\end{enumerate}

We use a BERT-like pretrained model $\mathcal{B}$ to encode the reference and candidate. 
In our experiments, $\mathcal{B}$ is a CodeBERT model that we further pretrained using the masked language modeling objective \citep{devlin2018bert} on language-specific corpora, but $\mathcal{B}$ can be any transformer-based model which we have access to its internal hidden states. 

\paragraph{Token Representation} We concatenate the context $x$ with each of the reference and the candidate, resulting in  $x\cdot y^*$ and $x\cdot \hat{y}$. We use the tokenizer $\mathcal{T}_\mathcal{B}$ provided with the model $\mathcal{B}$: 
\begin{equation}
    \begin{aligned}
    & \mathcal{T}_\mathcal{B}\left(x\cdot y^*\right) &&= \left<x_1,...,x_k,y^*_1,...,y^*_m\right> &\\
    &\mathcal{T}_\mathcal{B}\left(x\cdot \hat{y}\right) &&= \left<x_1,...,x_k,\hat{y}_1,...,\hat{y}_n\right> &
    \end{aligned}
\end{equation}
to get a sequences of tokens. We run a standard ``forward pass'' with the model $\mathcal{B}$ for each tokenized sequence, resulting in sequences of vectors: 
\begin{equation}
\begin{aligned}
    & \mathcal{B}\left(\left<x_1{,}...{,}x_k{,}y^*_1{,}...{,}y^*_m\right>\right){=}\left<\vx_1{,}...{,}\vx_k{,}\vy^*_1{,}...{,}\vy^*_m\right> & \\
    & \mathcal{B}\left(\left<x_1{,}...{,}x_k{,}\hat{y}_1{,}...{,}\hat{y}_n\right>\right) {=}\left<\vx_1{,}...{,}\vx_k{,}\hat{\vy}_1{,}...{,}\hat{\vy}_n\right> &
\end{aligned}\label{eq:forward}
\end{equation}
Finally, we mask out the encoded context tokens $\vx_1,...,\vx_k$ as well as all non-alphanumeric tokens (parentheses, brackets, dots, commas, whitespaces, etc.) except for arithmetic operators, from each of the encoded reference and encoded candidate. This results in encoded reference tokens $\vy^*=\left<\vy^*_1,...,\vy^*_m\right>$, encoded candidate tokens $\hat{\vy}=\left<\hat{\vy}_1,...,\hat{\vy}_n\right>$, and their corresponding masks $\vm^*$ and $\hat{\vm}$. We denote $\vy[\vm]$ as the remaining encoded tokens in $\vy$ after selecting only alphanumeric token vectors according to the mask $\vm$.

\paragraph{Similarity Computation} We compute the cosine similarity between the encoded reference and candidate tokens, following \citet{zhangbertscore}: 
\begin{equation}
    sim\left(y^*_i,\hat{y}_j \right)=
\frac{{\vy^*_i}^{\top} \cdot \hat{\vy}_j}{\norm{\vy^*_i}\cdot\norm{\hat{\vy}_j}}
\end{equation}
Although this compares the individual tokens $y^*_i$ and $\hat{y}_j$, their vector representations ${\vy^*_i}$ and $ \hat{\vy}_j$ contain information about their context, and thus about their semantic role in the code.

\paragraph{\cbs} We use the similarity matrix (see \Cref{fig:mainfigure}), formed by the similarity scores between $\vy^*$ and $\hat{\vy}$, to compute precision, recall, and F$_1$, by taking the maximum across the rows and columns of the similarity matrix, and then averaging. Following \citet{banerjee2005meteor}, we also compute F$_3$ by giving more weight to recall, as shown in \Cref{eq:main_eq}. Additional details regarding token weighting and scaling are provided in \Cref{sec:details}.


 

\section{Experimental Setup}


We evaluate \cbs across multiple datasets and programming languages. 
We first show that \cbs is more correlated with \emph{human preference} than previous metrics, using human-rated solutions for the CoNaLa dataset \citep{yin2018mining,evtikhiev2022out}. 
%
We then show that \cbs is more correlated with \emph{functional correctness}, using the HumanEval dataset \citep{chen2021evaluating}.
We also show that \cbs achieves a higher newly proposed \emph{distinguishability}  than other metrics (\Cref{sec:distinguishability}).
Finally, we analyze some of the design decisions and their implications.

\subsection{Training Language-specific CodeBERT models}
\paragraph{Training}
We used CodeBERT~\citep{feng2020codebert} 
as our base model ($\mathcal{B}$) and continued its self-supervised pretraining~\citep{gururangan2020don} with the masked language modeling (MLM) objective \citep{devlin2018bert} on Python, Java, C++, C, and JavaScript corpora. We trained a separate model for each programming language, for 1,000,000 steps for each language, using a batch size of 32, an initial learning rate of $5e^{-5}$, decayed linearly to $3e^{-5}$.
Our implementation is based on the widely used HuggingFace 
\texttt{Transformers} library \citep{wolf2019huggingface} and BERTScore\footnote{\url{https://github.com/Tiiiger/bert_score}}, and it supports any transformer-based model available on the HuggingFace hub.

\paragraph{Dataset} We trained each model on the language-specific subset of the CodeParrot \citep{tunstall2022natural} dataset\footnote{\url{https://huggingface.co/datasets/codeparrot/github-code-clean}}, which consists of overall 115M code files from GitHub, further filtered by keeping only files having average line length lower than 100, more than 25\% alphanumeric characters, and non-auto-generated files.
Even after 1,000,000 training steps, none of the models have completed even a single epoch, meaning that every training example was seen only once at most.


\subsection{Comparing Different Metrics}
We compare \cbs with existing metrics that are commonly used on code generation evaluation. 
We use human annotated preference and execution-based results as the ground truth and measure their correlation with these metrics.

\paragraph{Correlation metrics} We used three major correlation metrics. 
Following best practices in natural language evaluation, we used
\kt{}~(\ktt), \ps{}~(\pss) and \spe{}~(\spp) to measure the correlation between 
each metric's scores
and the references. 
The detailed equations can be found in Appendix \ref{sec:corr_metrics}.

\paragraph{Human preference experiments} 
We evaluate different metrics on CoNaLa~\citep{yin2018learning},
a natural language to Python code generation benchmark collected from StackOverflow. 
We use the human annotation of \citet{evtikhiev2022out} to measure the correlation between each metric and human preference. More details are provided in \Cref{app:eval_details_human}.

\paragraph{Functional correctness experiments} We evaluate functional correctness using the \he{}~\citep{chen2021evaluating} benchmark. 
Each example in \he contains a natural language goal, hand-written input-output test cases, and a human-written reference solution. 
While the original \he is in Python, \citet{cassano2022scalable} translated 
\he to 18 programming languages, and provided 
the predictions of the Codex model \citep{chen2021evaluating} (\davinci) and their corresponding functional correctness.\footnote{\url{https://huggingface.co/datasets/nuprl/MultiPL-E}}
We used Java, C++, Python, and JavaScript for these experiments, which are some of the most popular programming languages in open-source projects.\footnote{\url{https://octoverse.github.com/2022/top-programming-languages}}
More details are provided in \Cref{app:eval_details_functional}.


\paragraph{Hyperparameters} We tuned only the following hyperparameters for \cbs: whether to use F$_1$ or F$_3$, and which layer of the underlying model to extract the encoded tokens from, which we examine in \Cref{sec:analysis}.
%
We used F$_1$ in the human preference experiments and F$_3$ in the functional correctness experiments. 
We perform 3-fold cross-validation and report average results across the three folds.
As for the layer to extract the token vectors from, we used layer 7 for \conala, and in  \he{} we used layer 7 for Java, 10 for C++, 11 for JavaScript, and 9 for Python.


    \begin{table*}[t]
        \centering
        \begin{tabular}{lcccccccc}
        \toprule
         & \multicolumn{2}{c}{Java} & \multicolumn{2}{c}{C++} & \multicolumn{2}{c}{Python} & \multicolumn{2}{c}{JavaScript} \\
        Metric & \ktt & \spp & \ktt & \spp & \ktt & \spp & \ktt & \spp \\
        \midrule
        BLEU & .481 & .361 & .112 & .301 & .393 & .352 & .248 & .343 \\
    CodeBLEU & .496 & .324 & .175 & .201 & .366 & .326 & .261 & .299 \\
    ROUGE-1 & .516 & .318 & .262 & .260 & .368 & .334 & .279 & .280 \\
    ROUGE-2 & .525 & .315 & .270 & .273 & .365 & .322 & .261 & .292 \\
    ROUGE-L & .508 & .344 & .258 & .288 & .338 & .350 & .271 & .293 \\
    METEOR & \textbf{.558} & \textbf{.383} & .301 & .321 & .418 & .402 & \textbf{.324} & \textbf{.415} \\
    chrF & .532 & .319 & .319 & .321 & .394 & .379 & .302 & .374 \\
    CrystalBLEU & .471 & .273 & .046 & .095 & .391 & .309 & .118 & .059 \\
    \midrule
    \cbs & \textbf{.553} & .369 & \textbf{.327} & \textbf{.393} & \textbf{.422} & \textbf{.415} & \textbf{.319} & .402 \\
        \bottomrule
        \end{tabular}
        \caption{\kt (\ktt) and \spe (\spp) correlations of each metric with the functional correctness 
        on \he in multiple languages. 
        The correlation coefficients are reported as the average across three runs. Standard deviation is provided in \Cref{tab:corr_func_stdev_app}.}
        \label{tab:corr_func_stdev}
    \end{table*}
\begin{table}
    \centering
    \begin{tabular}{lccc}
    \toprule
    Metric & \ktt & \pss & \spp \\
    \midrule
    BLEU & .374 & .604 & .543 \\
    CodeBLEU & .350 & .539 & .495 \\
    ROUGE-1 & .397 & .604 & .570 \\
    ROUGE-2 & .429 & .629 & .588 \\
    ROUGE-L & .420 & .619 & .574 \\
    METEOR & .366 & .581 & .540 \\
    chrF & .470 & .635 & .623 \\
    CrystalBLEU & .411 & .598 & .576 \\
    \midrule
    CodeBertScore & \textbf{.517} & \textbf{.674} & \textbf{.662} \\
    \bottomrule
    \end{tabular}
    \caption{The \kt{} (\ktt), \ps{} (\pss) and \spe{} (\spp) correlation with human preference. 
    The best performance is \textbf{bold}.
    The correlation coefficients are reported as the average across three runs. Standard deviations are provided in \Cref{tab:corr_human_stdev_app}. }
    \label{tab:corr_human_stdev}
\end{table}

\section{Results}
\label{sec:results}
\paragraph{Correlation with human preference} \Cref{tab:corr_human_stdev} shows the correlation between different metrics and human preference. 
\cbs achieves the highest correlation with human preference, across all correlation metrics.
While \citet{evtikhiev2022out} suggested that chrF and ROUGE-L are the most suitable metrics for evaluating code generation models in \conala,
\cbs outperforms these metrics by a significant margin.
For example, \cbs achieves \kt correlation 
of 0.517 compared to 0.470 of chrF and 0.420 of ROUGE-L.
These results show that  generated code that is preferred by \cbs --- also tends to be preferred by human programmers.



\paragraph{Correlation with functional correctness} \Cref{tab:corr_func_stdev} shows the correlation between different metrics and functional correctness:
\cbs  achieves the highest or comparable \kt and \spe correlation with  functional correctness across \emph{all four} languages.
METEOR  achieves a comparable  correlation with \cbs in Java and JavaScript, and its correlation is surprisingly better than other baseline metrics. However, in C++ and Python, \cbs is strictly better.
Overall on average across languages, \cbs is more correlated with functional correctness than all baselines.

\section{Analysis}\label{sec:analysis}

\begin{figure*}[t]
    \definecolor{nicegreen}{HTML}{0F9D58}
    \definecolor{niceorange}{HTML}{F8BA02}
\centering
\small
\begin{tikzpicture}
\begin{axis}[
ybar,
ymin=0,
ymax=0.8,
x=3cm,
bar width=0.5cm,
height=6cm,
width=1.05\textwidth,
enlarge x limits=0.15,
xtick={1,2,3,4,5},
xticklabels={CoNaLa, HumanEval-Python, HumanEval-Java, HumanEval-C++, HumanEval-JavaScript},
ytick={0,0.1,0.2,0.3,0.4,0.5,0.6,0.7},
              legend style={at={(0.55,1)},anchor=north,
              mark size=2pt,
                /tikz/every even column/.append style={column sep=2mm},
                legend columns=-1
                },
              legend cell align={left},
nodes near coords,
nodes near coords style={font=\tiny},
]
\addplot[black, fill=red] coordinates {(1,0.5241)(2,0.4563)(3,0.5574)(4,0.3844)(5,0.3710)};
\addplot[black, fill=blue] coordinates {(1,0.5239)(2,0.4459)(3,0.5571)(4,0.3939)(5,0.3843)};
\addplot[black, fill=niceorange] coordinates {(1,0.6510)(2,0.3383)(3,0.3257)(4,0.3741)(5,0.3441)};
\addplot[black, fill=nicegreen] coordinates {(1,0.6503)(2,0.3492)(3,0.3228)(4,0.3488)(5,0.3044)};
\legend{\ktt{}-Lang-specific, \ktt{}-Base model, \spp{}-Lang-specific, \spp{}-Base model}
\end{axis}
\end{tikzpicture}
\caption{The \kt and \spe on the development set of different datasets with the language-specific pretrained model~(Lang-specific) and with the base  CodeBERT~(Base model). 
}
\label{fig:lang_base_compare}
\end{figure*}

\begin{figure}[t!]
\centering
    \begin{tikzpicture}
\begin{axis}[
xmin=0, xmax=13,
ymin=0.25, ymax=0.45,
xlabel={Layer},
legend pos=south east,
              legend style={at={(1,0)},anchor=south east,
              mark size=2pt,
                /tikz/every even column/.append style={column sep=2mm},
                },
              legend cell align={left},
height=7cm,
width=1\linewidth,
]
\addplot[
    color=blue,
    mark=square*,
    mark size=2pt,
    line width=2pt,
    mark options={solid, fill=blue, draw=black, line width=1pt},
    ] coordinates {
(1,0.3743)(2,0.3939)(3,0.406)(4,0.4115)(5,0.4213)(6,0.4267)(7,0.4347)(8,0.4416)(9,0.4372)(10,0.4386)(11,0.4409)(12,0.4344)
};

\addplot[
    color=red,
    mark=triangle*,
    line width=2pt,
    mark size=2pt,
    mark options={solid, fill=red, draw=black, line width=1pt},
    ] coordinates {
(1,0.295)(2,0.3302)(3,0.3506)(4,0.3629)(5,0.3668)(6,0.3706)(7,0.3767)(8,0.374)(9,0.3734)(10,0.3758)(11,0.3793)(12,0.3747)
};

\addplot[
    color=green!70!black,
    mark=*,
    line width=2pt,
    mark size=2pt,
    mark options={solid, fill=green!70!black, draw=black, line width=1pt},
    ] coordinates {
(1,0.3322)(2,0.3293)(3,0.3307)(4,0.3365)(5,0.3451)(6,0.3407)(7,0.3462)(8,0.357)(9,0.3558)(10,0.3563)(11,0.3562)(12,0.3575)
};

\addplot[
    color=yellow!80!black,
    mark=diamond*,
    line width=2pt,
    mark size=2pt,
    mark options={solid, fill=yellow!80!black, draw=black, line width=1pt},
    ] coordinates {
(1,0.3092)(2,0.3301)(3,0.3447)(4,0.3605)(5,0.378)(6,0.3879)(7,0.3921)(8,0.3952)(9,0.3958)(10,0.3973)(11,0.3931)(12,0.3855)
};

\legend{Java, C++, JavaScript, Python}
\end{axis}
\end{tikzpicture}
\caption{The average of \kt and \spe on the development set of \he when using the embeddings from different layers. }
\label{fig:layers}
\end{figure}
\begin{figure*}[t!]
    \newcommand{\heatmapheight}{6cm}
    \centering
    \begin{minipage}{0.4\textwidth}
    \centering
    \begin{subfigure}{1\textwidth}
      \centering      \includegraphics[height=\heatmapheight]{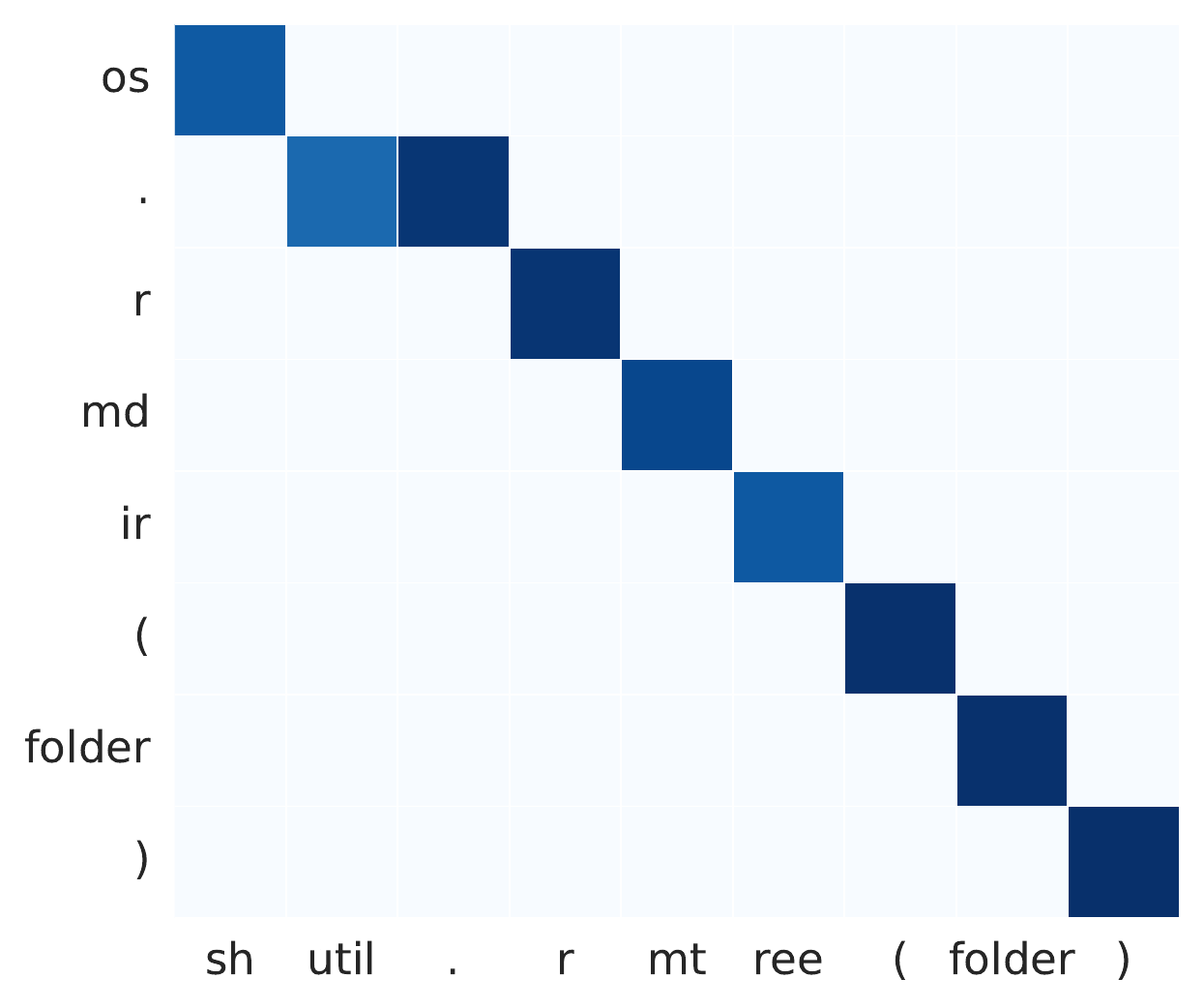}
      \caption{}
      \label{fig:sub1}
    \end{subfigure}%
    \end{minipage}
    \hspace{0.05\textwidth}
    \begin{minipage}{0.4\textwidth}
    \centering
    \begin{subfigure}{1\textwidth}
      \centering
      \includegraphics[height=\heatmapheight]{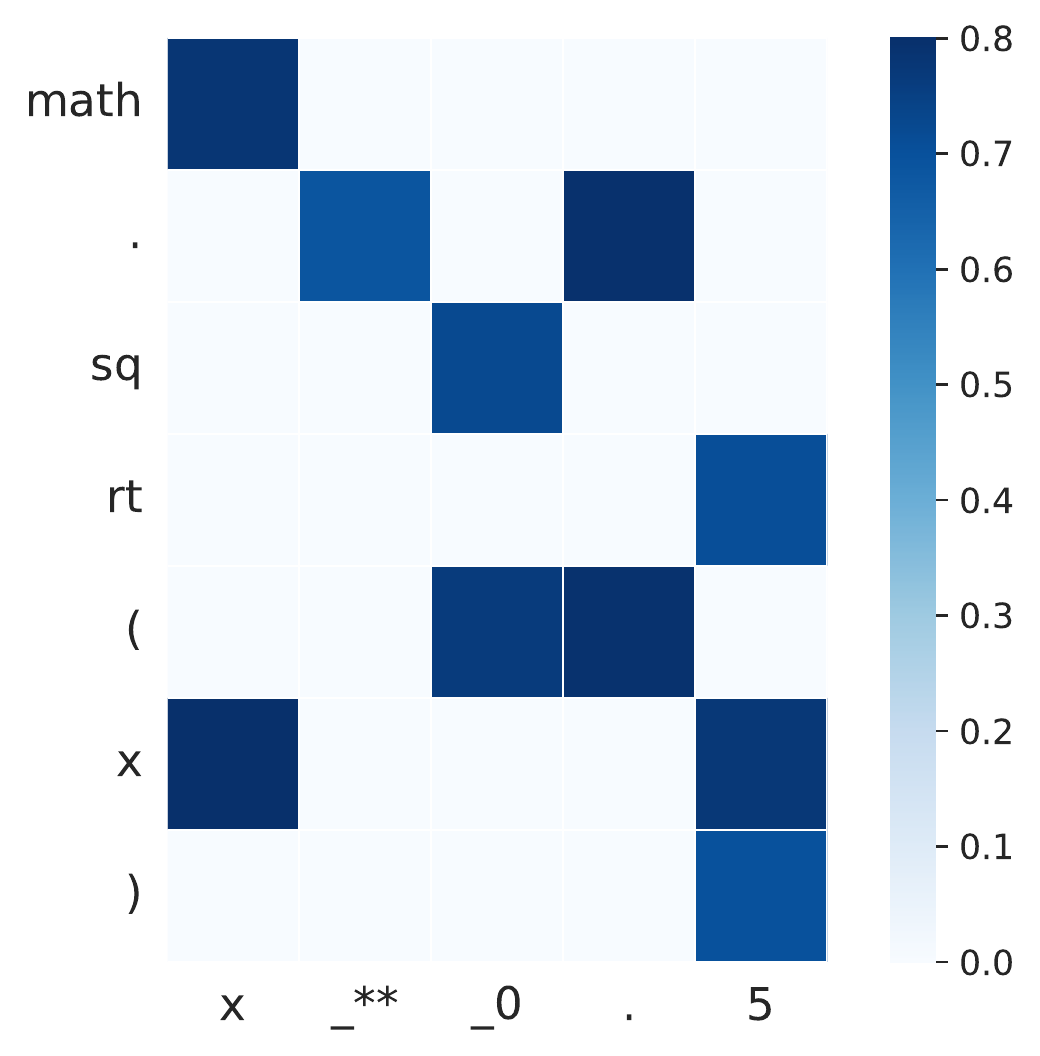}
      \caption{}
      \label{fig:sub2}
    \end{subfigure}
    \end{minipage}
    \caption{Heatmaps of the similarity scores between two  pieces of code that achieve the same goal. \Cref{fig:sub1} shows the similarity scores  between \texttt{os.rmdir(folder)} and \texttt{shutil.rmtree(folder)}. 
    \Cref{fig:sub2} shows the similarity scores between \texttt{math.sqrt(x)} and \texttt{x ** 0.5}. 
    }
    \label{fig:heatmap}
\end{figure*}

We conducted a series of additional experiments to understand the importance of different design decisions, and to gain insights on applying \cbs to new datasets and scenarios.

\paragraph{Can we use \cbs in a new language without a language-specific CodeBERT?}
In all experiments in \Cref{sec:results}, we used the language-specific model which we continued to pretrain on each language. 
But what if we wish to use \cbs in a language in which we don't have a language-specific model?
We compare the language-specific models to CodeBERT-base in
\Cref{fig:lang_base_compare}.
Generally,  CodeBERT-base achieves close performance to a language-specific model. 
However, in most \he experiments and correlation metrics, using the language-specific model \emph{is} beneficial.
These results show that language-specific models are often preferred if such models are available, but
the CodeBERT-base can still provide close performance even without language-specific pretraining.

\paragraph{Which transformer layer should we use?}
We further investigate the impact of using hidden states from different layers of the model --- the layer which the vectors in \Cref{eq:forward} come from, in the computation of \cbs. 
The results are shown in \Cref{fig:layers}:
generally, the deeper the layer -- the higher the average correlation between \cbs and functional correctness, across all programming languages.
However in almost all languages, performance reaches its maximum before the last layer, and decreases at the following layers.
This suggests that higher layers encode the semantic information of each token more accurately, but the final layers may be more task-specific.
These observations are consistent with
\citet{tenney2019bert}, who found that lower layers in BERT tend to process shallow information, while higher layers encode deeper semantic meaning in natural language.

\paragraph{Does encoding natural language context help?}
One major difference between \cbs and BERTScore is that \cbs leverages the \emph{context} for the generated code, such as the natural language instruction or intent that was given as input for generation. 
We find that using context increases the correlation, for example, the \kt of \cbs from 0.50 to 0.52. 
While this paper mainly focuses on natural language instructions, we believe that \cbs can thus benefit other programming scenarios as well, for example when generating code given the human-written comments, or generating code given the preceding code context.

\paragraph{\cbs allows soft matching of tokens} The heatmaps in \Cref{fig:heatmap} show the similarity scores between tokens in \cbs. 
For example, both \texttt{shutil.rmtree} and \texttt{os.rmdir} in \Cref{fig:sub1} delete a \texttt{folder}; \cbs aligns each token to a respective token in the other expression, even though the two spans do not share many identical tokens. 

In \Cref{fig:sub2}, both code snippets calculate a square root, 
where one uses \texttt{math.sqrt(x)} and the other uses \texttt{x ** 0.5}. 
An exact surface-form-matching metric such as chrF
would assign a low similarity score to this code pair, as they only share the token \texttt{x}. However, \cbs assigns non-zero scores to each token with meaningful alignments, such as matching \texttt{[sq,rt]} with \texttt{[\_0,5]}, since a square root is the 0.5-th power.

Additionally, we study the robustness of \cbs to adversarial perturbations. We found that token-based metrics such as chrF are much more prone to matching trivial tokens rather than tokens that preserve the semantic meaning of the code. Examples can be found in \Cref{sec:robustness}.

Additional discussion and experiments regarding the distinguishability of \cbs are provided in \Cref{sec:distinguishability}.
Additional general examples are provided in \Cref{sec:additional_examples}.

\section{Related Work}
\paragraph{Token-based metrics} Metrics such as BLEU \citep{papineni2002bleu} evaluate code generation by counting matching n-grams between generated and reference code. CrystalBLEU \citep{eghbali2022crystalbleu} refines this approach by disregarding trivially shared n-grams, while ROUGE \citep{lin2004rouge} and METEOR \citep{banerjee2005meteor} emphasize recall and balance of precision and recall respectively. However, these metrics, relying on \emph{exact} lexical matches, often fail to capture semantically equivalent but lexically different code snippets. Unlike these, \cbs captures the wide, two-sided \emph{context} of each token, which n-grams cannot capture.

\paragraph{Static analysis-based metrics}
CodeBLEU \citep{ren2020codebleu} incorporates data-flow and Abstract Syntax Tree (AST) matching, in addition to token-matching.
However, valid code may not always align in ASTs and data-flows. 
Additionally, partial code, although potentially useful, may not parse, thus cannot be fully evaluated by CodeBLEU. 
Further, as highlighted by subsequent studies \citep{wang2022execution}, CodeBLEU does not correlate well with execution accuracy.

\paragraph{Execution-based Metrics}
To alleviate previous issues, execution-based evaluation counts a generated code snippet as correct if it produces the required outputs when run with given inputs \citep{chen2021evaluating,athiwaratkun2022multi,li2022competition,wang2022execution,lai2022ds,huang2022execution}.
However, execution-based evaluation requires datasets that are provided with manually crafted test cases for each example, which is costly and labor-intensive to create; thus, only few such datasets exist. 
In contrast, \cbs is completely unsupervised and does not depend on any specific dataset.
Further, executing model-generated code is susceptible to security threats, and thus should be run in an isolated sandbox, which makes it technically cumbersome to work with iteratively.



\section{Conclusion}
In this paper, we present \cbs, a simple evaluation metric for code generation, which builds on BERTScore \citep{zhangbertscore}, using pretrained language models of code, and leveraging the natural language context of the generated code. We perform an extensive evaluation across four programming languages which shows that \cbs is more correlated with human preference than all prior metrics. Further, we show that generated code that receives a higher score by \cbs is more likely to function correctly when executed. Finally, we release five programming language-specific pretrained models to use with our publicly available code. These models were downloaded more than \downloaded times from the HuggingFace Hub. Our code and data are available at \url{https://github.com/neulab/code-bert-score}.

\section*{Acknowledgement}
We thank Misha Evtikhiev, Egor Bogomolov, and Timofey Bryksin for the discussions, and for the data from their paper \citep{evtikhiev2022out}.
We thank anonymous reviewers for the valuable feedback.
We are grateful to Yiwei Qin for the discussions regarding the T5Score paper \citep{qin2022t5score}; the idea to use functional correctness as a meta-metric was born thanks to the discussion with her.
We are also grateful to Aryaz Eghbali and Michael Pradel for the discussions about CrystalBLEU \citep{eghbali2022crystalbleu}. 
This material is partly based on research sponsored in part by the Air Force Research Laboratory under agreement number FA8750-19-2-0200. The U.S. Government is authorized to reproduce and distribute reprints for Governmental purposes notwithstanding any copyright notation thereon. The views and conclusions contained herein are those of the authors and should not be interpreted as necessarily representing the official policies or endorsements, either expressed or implied, of the Air Force Research Laboratory or the U.S. Government. This project was also partially supported by a gift from AWS AI.

\section*{Limitations}
\cbs requires a GPU for computing the metric, while traditional metrics such as BLEU require only a CPU. 
This adds a hardware requirement to the evaluation of models of code, while most previous approaches are computationally cheaper (\eg by counting n-grams).
However, since training and testing neural models require GPU anyways, we can safely assume that a GPU is available. 
Further, BERT-base models are encoder-only and non-autoregressive; this means that they require only a \emph{single} ``forward pass'', compared to encoder-decoder models (\eg T5) and decoder-only models (\eg GPT-3) that need to autoregressively generate token after token, using a forward pass for each output token. Thus, 
the additional time consumption by encoder-only models~(\eg BERT) is negligible, especially when evaluating encoder-decoder or decoder-only as the \nlcode generator models.

Another point to consider is that \cbs relies on a strong underlying BERT-based model, while methods such as BLEU do not have many ``moving parts'' or hyperparameters to tune. 
However, this is mostly an advantage, since \cbs can be further improved in the future using stronger base models.

\bibliographystyle{acl_natbib}
\bibliography{references}

\appendix

\clearpage
\section{Additional Details}
\label{sec:details}

\paragraph{F$_\beta$} The well-known F$_1$ score is computed as: 
\begin{equation}
    F_1 = \frac{2}{\frac{1}{\textrm{recall}}+\frac{1}{\textrm{precision}}}=\frac{2\cdot \textrm{precision} \cdot \textrm{recall}}{\textrm{precision} + \textrm{recall}}
\end{equation}
A more general F score $F_{\beta}$ uses a positive factor $\beta$, where recall is considered $\beta$ times as important as precision:
\begin{equation}
    F_{\beta} = \frac{\left(1+\beta^2\right)\cdot \textrm{precision} \cdot \textrm{recall}}{\beta^2\cdot \textrm{precision} + \textrm{recall}}
\end{equation}
As found in METEOR \citep{banerjee2005meteor}, using F$_\beta$ with $\beta=3$, thus preferring recall over precision, results in a higher correlation with human preference in machine translation. In our experiments, we found that this applies to \nlcode as well.

\paragraph{Token Weighting} Following \citet{zhangbertscore}, we compute the inverse document frequency (idf), according to a language-specific test set, and weigh each token according to its negative log frequency. 

\paragraph{Scaling} Following \citet{zhangbertscore}, the cosine similarity scores of hidden states tend to lie in a limited range. Thus, we can linearly scale the resulting scores, using an empirical base scalar $b$:
\begin{equation}
    \widehat{\mathrm{CodeBERTScore}} = \frac{\mathrm{CodeBERTScore} - b}{1 - b}
\end{equation}
This typically spreads the \cbs F$_1$ scores to the $[0,1]$ range, and is merely a cosmetical change: this scaling does not change the way \cbs \emph{ranks} different prediction, but can be slightly more intuitive and easier to interpret.
We computed $b$ empirically by sampling random unrelated code pairs and measuring their average similarity score.
For Java, the empirical $b_{\mathrm{Java}}$ was 0.78 and for C++,  $b_{\mathrm{C++}}$ it was 0.76.

\section{Evaluation Details}
\label{app:eval_details}

\subsection{Human Preference}\label{app:eval_details_human}
For each example, \citet{evtikhiev2022out} asked experienced software developers to grade the generated code snippets from five different models. 
The grade scales from zero to four, with zero denoting that the generated code is irrelevant and unhelpful, and four meaning that the generated code solves the problem accurately. 
Overall, there are 2860 annotated code snippets (5 generations $\times$ 472 examples) where each snippet is graded by 4.5 annotators. 

\subsection{Functional Correctness}\label{app:eval_details_functional}
We evaluate functional correctness using the \he{}~\citep{chen2021evaluating} benchmark. 
Each example in \he contains a natural language goal, hand-written input-output test cases, and a human-written reference solution. 
On average, each example has 7.7 test cases and there are 164 examples in total.
While the original \he is in Python, \citet{cassano2022scalable} translated 
\he to 18 programming languages, and provided 
the predictions of the Codex model \citep{chen2021evaluating} (\davinci) and their corresponding functional correctness.\footnote{\url{https://huggingface.co/datasets/nuprl/MultiPL-E}}
We used Java, C++, Python, and JavaScript for these experiments, which are some of the most popular programming languages in open-source projects.\footnote{\url{https://octoverse.github.com/2022/top-programming-languages}}
Notably, \citet{cassano2022scalable} did not translate the reference solutions to the other languages, so, we collected these from \he{}-X~\citep{zeng2022glm}.\footnote{\url{https://huggingface.co/datasets/THUDM/humaneval-x}} 
The reference score of every example is either 1 (``correct'', if it passes all test cases) or 0 (``incorrect'', otherwise).

\section{Correlation Metrics}\label{sec:corr_metrics}
\paragraph{\kt{}~(\ktt)} \ktt measures the \emph{ordinal/rank} association between a metric such as \cbs and the reference measurement. It is calculated as:
\begin{equation*}
    \tau = \frac{|\text{concordant}| - |\text{discordant}|}{|\text{concordant}| + |\text{discordant}|}
\end{equation*}
where $|\text{concordant}|$ represents the number of pairs where two measurements agree on their relative rank. That is, if $f(\hat{y_1}, y_1^*) > f(\hat{y_2}, y_2^*)$, the reference measurement also yields $\fref(\hat{y_1}, y_1^*) > \fref(\hat{y_2}, c_2^*)$.
Similarly, $|\text{discordant}|$ represents the number of pairs where two measurements yield opposite ranks.
Notably, in our experiments, we restrict the comparisons of ranks within the generations of the \emph{same} question. 

\paragraph{\ps{}~(\pss{})} \pss measures the \emph{linear} correlation between a metric and the reference measurement. It is defined as:
\begin{equation*}
\resizebox{\columnwidth}{!}{%
    $r_p = \frac{\sum_{i=1}^N(f(\hat{y_i}, y_i^*) - \bar{f})(\fref(\hat{y_i}, y_i^*) - \bar{\fref})}{\sqrt{\sum_{i=1}^N(f(\hat{y_i}, y_i^*) - \bar{f})^2\sum_{i=1}^N(\fref(\hat{y_i}, y_i^*) - \bar{\fref})^2}}$
}
\end{equation*}

where $N$ is the number of generations in the dataset, $\bar{f}$ is the mean \cbs of the dataset, and $\bar{\fref}$ is the mean similarity score calculated by the reference measurement.  

\paragraph{\spe{}~(\spp)}
\spp measures the Pearson correlation coefficient between the \emph{ranks} produced by a metric and the reference measurement:

\begin{equation*}
    r_p = \frac{\textrm{cov}(R(f(\hat{\mathbf{Y}}), R(\fref(\mathbf{Y}^*)))}{\sigma_{R(f(\hat{\mathbf{Y}))}}
    \sigma_{R(\fref(\mathbf{Y}^{*}))}}
\end{equation*}
where $R$ returns the ranks of  code snippets in a collection of code snippets $\mathbf{Y}$. $\textrm{cov}(\cdot, \cdot)$ is the covariance of two variables and $\sigma(\cdot)$ is the standard deviation. 

\section{Standard Deviation}
\Cref{tab:corr_func_stdev_app} shows the same results as in \Cref{tab:corr_func_stdev}, but with standard deviations.
\Cref{tab:corr_human_stdev_app} shows the results from \Cref{tab:corr_human_stdev}, with standard deviations.

\begin{table*}[t]
    \centering
    \footnotesize
    \resizebox{1\textwidth}{!}{ 
    \begin{tabular}{lcccccccc}
    \toprule
     & \multicolumn{2}{c}{Java} & \multicolumn{2}{c}{C++} & \multicolumn{2}{c}{Python} & \multicolumn{2}{c}{JavaScript} \\
    Metric & \ktt & \spp & \ktt & \spp & \ktt & \spp & \ktt & \spp \\
    \midrule
    BLEU & .481\scriptsize($\pm$.030) & .361\scriptsize($\pm$.037) & .112\scriptsize($\pm$.059) & .301\scriptsize($\pm$.054) & .393\scriptsize($\pm$.083) & .352\scriptsize($\pm$.064) & .248\scriptsize($\pm$.075) & .343\scriptsize($\pm$.052) \\
CodeBLEU & .496\scriptsize($\pm$.034) & .324\scriptsize($\pm$.037) & .175\scriptsize($\pm$.021) & .201\scriptsize($\pm$.037) & .366\scriptsize($\pm$.079) & .326\scriptsize($\pm$.075) & .261\scriptsize($\pm$.065) & .299\scriptsize($\pm$.043) \\
ROUGE-1 & .516\scriptsize($\pm$.052) & .318\scriptsize($\pm$.043) & .262\scriptsize($\pm$.073) & .260\scriptsize($\pm$.024) & .368\scriptsize($\pm$.092) & .334\scriptsize($\pm$.054) & .279\scriptsize($\pm$.092) & .280\scriptsize($\pm$.068) \\
ROUGE-2 & .525\scriptsize($\pm$.049) & .315\scriptsize($\pm$.047) & .270\scriptsize($\pm$.073) & .273\scriptsize($\pm$.036) & .365\scriptsize($\pm$.094) & .322\scriptsize($\pm$.077) & .261\scriptsize($\pm$.077) & .292\scriptsize($\pm$.057) \\
ROUGE-L & .508\scriptsize($\pm$.060) & .344\scriptsize($\pm$.038) & .258\scriptsize($\pm$.091) & .288\scriptsize($\pm$.027) & .338\scriptsize($\pm$.103) & .350\scriptsize($\pm$.064) & .271\scriptsize($\pm$.078) & .293\scriptsize($\pm$.046) \\
METEOR & \textbf{.558}\scriptsize($\pm$.058) & \textbf{.383}\scriptsize($\pm$.027) & .301\scriptsize($\pm$.061) & .321\scriptsize($\pm$.023) & .418\scriptsize($\pm$.090) & .402\scriptsize($\pm$.049) & \textbf{.324}\scriptsize($\pm$.075) & \textbf{.415}\scriptsize($\pm$.022) \\
chrF & .532\scriptsize($\pm$.067) & .319\scriptsize($\pm$.035) & .319\scriptsize($\pm$.056) & .321\scriptsize($\pm$.020) & .394\scriptsize($\pm$.096) & .379\scriptsize($\pm$.058) & .302\scriptsize($\pm$.073) & .374\scriptsize($\pm$.044) \\
CrystalBLEU & .471\scriptsize($\pm$.024) & .273\scriptsize($\pm$.067) & .046\scriptsize($\pm$.009) & .095\scriptsize($\pm$.064) & .391\scriptsize($\pm$.080) & .309\scriptsize($\pm$.073) & .118\scriptsize($\pm$.057) & .059\scriptsize($\pm$.069) \\
\midrule
\cbs & \textbf{.553}\scriptsize($\pm$.068) & .369\scriptsize($\pm$.049) & \textbf{.327}\scriptsize($\pm$.086) & \textbf{.393}\scriptsize($\pm$.048) & \textbf{.422}\scriptsize($\pm$.090) & \textbf{.415}\scriptsize($\pm$.071) & \textbf{.319}\scriptsize($\pm$.054) & .402\scriptsize($\pm$.030) \\
    \bottomrule
    \end{tabular}
    }
    \caption{\kt (\ktt) and \spe (\spp) correlations of each metric with the functional correctness 
    on \he in multiple languages. 
    The correlation coefficients are reported as the average across three runs, along with the standard deviation.}
    \label{tab:corr_func_stdev_app}
\end{table*}

\begin{table*}
    \centering\small
    \begin{tabular}{lccc}
    \toprule
    Metric & \ktt & \pss & \spp \\
    \midrule
    BLEU & .374\scriptsize($\pm$.025) & .604\scriptsize($\pm$.016) & .543\scriptsize($\pm$.018) \\
    CodeBLEU & .350\scriptsize($\pm$.037) & .539\scriptsize($\pm$.033) & .495\scriptsize($\pm$.037) \\
    ROUGE-1 & .397\scriptsize($\pm$.023) & .604\scriptsize($\pm$.016) & .570\scriptsize($\pm$.018) \\
    ROUGE-2 & .429\scriptsize($\pm$.025) & .629\scriptsize($\pm$.015) & .588\scriptsize($\pm$.022) \\
    ROUGE-L & .420\scriptsize($\pm$.037) & .619\scriptsize($\pm$.014) & .574\scriptsize($\pm$.022) \\
    METEOR & .366\scriptsize($\pm$.033) & .581\scriptsize($\pm$.016) & .540\scriptsize($\pm$.022) \\
    chrF & .470\scriptsize($\pm$.029) & .635\scriptsize($\pm$.023) & .623\scriptsize($\pm$.018) \\
    CrystalBLEU & .411\scriptsize($\pm$.030) & .598\scriptsize($\pm$.019) & .576\scriptsize($\pm$.034) \\
    \midrule
    CodeBertScore & \textbf{.517}\scriptsize($\pm$.024) & \textbf{.674}\scriptsize($\pm$.012) & \textbf{.662}\scriptsize($\pm$.012) \\
    \bottomrule
    \end{tabular}
    \caption{The \kt{} (\ktt), \ps{} (\pss) and \spe{} (\spp) correlation with human preference. 
    The best performance is \textbf{bold}.
    The correlation coefficients are reported as the average across three runs. Numbers inside parentheses indicate the standard deviations. }
    \label{tab:corr_human_stdev_app}
\end{table*}

\section{Robustness to adversarial perturbations}\label{sec:robustness}
\begin{figure}[h!]
\small
\begin{tabular}{lrr}
    \multicolumn{3}{l}{\underline{Ref}: \texttt{shutil.rmtree(folder)}} \\ \\
       \toprule
       Candidate & \cbs & chrF \\
       \midrule
       \textmd{\texttt{os.rmdir(folder)}} & 1st & 1st \\
       \textmd{\texttt{os.rmdir(f)}} & 2nd & 3rd \\
       \textmd{\texttt{(folder)}} & 3rd & 2nd \\
       \bottomrule
    \end{tabular}
    \caption{The similarity rankings of three code snippets given the reference code \texttt{shutil.rmtree(folder)}. While  \cbs correctly ranks \texttt{os.rmdir(f)} over the the non-equivalent \texttt{(folder)}, chrF prefers just \texttt{(folder)} over \texttt{os.rmdir(f)}.}
    \label{tab:pertub}
\end{figure}
We conducted a qualitative evaluation of \cbs under various perturbations.
An example is shown in \Cref{tab:pertub}, which shows the \cbs and chrF rankings of three code snippets based on the similarity to the reference \texttt{shutil.rmtree(folder)}. 
\cbs gives a higher ranking to the code snippet that employs the appropriate API~(\texttt{os.rmdir}) than the trivial \texttt{(folder)} that has the same variable name but without any function call. 
Contrarily, chrF assigns a higher ranking to \texttt{(folder)} which has a longer common sequence of characters, although semantically inequivalent.

\section{Distinguishing Code with Different Semantics}
\label{sec:distinguishability}
We study how well can \cbs perform as a generic similarity function that measures the similarity between two arbitrary code snippets $y_i$ and $y_j$.

\subsection{Distinguishability Metric}
We evaluate \cbs 
using the distinguishability metric $d$ proposed by \citet{eghbali2022crystalbleu} which is calculated as follows:
\begin{equation}
    d = \frac{\sum_{y_i, y_j \in \text{Pairs}_{\text{intra}}} f(y_i, y_j)}{\sum_{y_i, y_j \in \text{Pairs}_{\text{inter}}} f(y_i, y_j)}
    \label{eq:d}
\end{equation}
where $\text{Pair}_{\text{intra}}$ defines a set of code pairs from the same semantically equivalent clusters, and $\text{Pair}_{\text{inter}}$ defines a set of code pairs from two clusters of different functionality. Formally,
\begin{equation*}
    \begin{aligned}
        &\text{Pair}_{\text{intra}} &=& \{(y_i, y_j) \mid \exists k \textrm{ such that }  y_i, y_j \in C_k\} \\
        &\text{Pair}_{\text{inter}} &=& \{(y_i, y_j) \mid \exists k \textrm{ such that } y_i \in C_k, y_j \notin C_k\}
    \end{aligned}
\end{equation*}
where $C_k$ is the $k$-th cluster with semantically equivalent code snippets.
Intuitively, a similarity function $f$ that can distinguish between similar and dissimilar code will produce $d$ larger than 1, meaning that a pair of code snippets from the same semantic cluster has a higher similarity score than a pair of snippets from different clusters.
Since the number of intra-class and inter-class pairs grows quadratically with the number of code snippets, in our experiments we followed \citet{eghbali2022crystalbleu} to sample $N$ inter- and $N$ intra-class pairs instead. 

\subsection{Dataset with Semantically equivalent clusters} 
We follow~\citet{eghbali2022crystalbleu} to evaluate whether \cbs can distinguish similar and dissimilar code mined from \share\footnote{\url{https://sharecode.io/}}, an online coding competition platform. 
Semantically equivalent code snippets are from the same coding problem, and they all pass the unit tests provided by the platform. 
The dataset consists 6958 code snippets covering 278 problems in Java and C++. 
We use \cbs to calculate the similarity score for code pairs that share the same semantic class and code pairs that do not.
We then measure the distinguishability of \cbs according to \autoref{eq:d}. The results are shown in \autoref{tab:distinguish}.


\begin{table}[t!]
\centering\small
    \begin{tabular}{lcc}
    \toprule
    Metric & Java & C++ \\
    \midrule
    BLEU & 2.36 & 2.51\\
    CodeBLEU  & 1.44 & 1.42\\
    CrystalBLEU & 5.96 & 6.94\\
    \midrule
    \cbs & \textbf{9.56} & \textbf{9.13}\\
    \bottomrule
    \end{tabular}
    \caption{Distinguishability with different metrics as the similarity function. \cbs achieves a higher distinguishability than CrystalBLEU, which proposed this meta-metric, on the same datasets.
    }
    \label{tab:distinguish}
\end{table}

\Cref{tab:distinguish} shows that \cbs achieves a higher \emph{distinguishability} than CrystalBLEU, which proposed this meta-metric, in both Java and C++. \cbs achieves distinguishability scores of 9.56 in Java while CrystalBLEU achieves 5.96; in C++, \cbs achieves 9.13 while CrystalBLEU achieves only 6.94.
This result confirms that \cbs assigns higher similarity scores to semantically similar code pairs, compared to randomly paired snippets that belong to different semantic classes.





\paragraph{Can We Hack the Distinguishability Metric?}
Despite the encouraging results in \Cref{tab:distinguish}, we also found that distinguishability 
can be easily manipulated since it compares \emph{absolute} scores across different metrics. For example, while CrystalBLEU achieves a distinguishability score of 5.96, we can craft a variant of \cbs that achieves a distinguishability score of 120,000 by simple exponentiation of \cbs's output score. 


To illustrate this, we conducted a distinguishability evaluation with the same configurations as before, but with a variant of \cbs that we call CodeBERTScore$^k$, and defined as the composition of \cbs with the $f\left(x\right)=x^k$ function, that is: CodeBERTScore$^k\left(y_1,y_2\right)=\left(\mathrm{CodeBERTScore}\left(y_1,y_2\right)\right)^k$.




\begin{figure}[h!]
    \pgfplotsset{scaled y ticks=false}
    \begin{tikzpicture}
    \begin{axis}[
        xlabel=k,
        xmin=0, xmax=51,
        ymin=0, ymax=130000,
        xtick={0, 10, 20, 30, 40, 50},
        ytick={0,10000,20000,30000,40000,50000,60000,70000,80000,90000,100000,110000,120000,130000,140000,150000},
        legend pos=north west,
        yticklabel style={
            /pgf/number format/fixed,
            /pgf/number format/precision=0
        },
        grid = major,
        major grid style={dotted,gray},
        height=7cm,
        width=1\linewidth,
    ]
    
    \addplot[
        color=blue,
        mark=square*,
        line width=1pt,
        mark options={solid, fill=blue, draw=black},
        ]
        coordinates {
        (1, 9.55430033257913)
        (10, 61.38513021578588)
        (20, 1988.187107227889)
        (30, 8130.883552581966)
        (40, 31983)
        (50, 123319)
        };
        \addlegendentry{\cbs$^k$}
    
    \end{axis}
    \end{tikzpicture}
    \caption{Distinguishability by exponentiating the original \cbs by $k$.%
    }\label{fig:dist_hack}
    \end{figure}

As \autoref{fig:dist_hack} shows, distinguishability of CodeBERTScore$^k$ increases almost exponentially while increasing  $k$, although the base \cbs metric has not changed.

We thus argue that distinguishability is not a reliable meta-metric and is no substitute for execution-based- or human-rating. 
We further suspect that any meta-metric that compares exact, absolute, scores across different metrics is susceptible to such manipulations, and the reliable way to compare metrics is according to the way they \emph{rank} different examples, rather than the exact scores.


The distinguishability results of CodeBERTScore$^k$ with different values of $k$ are shown in \autoref{fig:dist_hack}. 
As \autoref{fig:dist_hack} shows, the distinguishability increases almost exponentially with the increasing value of $k$. 
We thus argue that distinguishability is not a reliable meta-metric and is no substitute for execution-based- or human-rating. 
We further suspect that any meta-metric that compares exact, absolute, scores across different metrics is susceptible to such manipulations, and the reliable way to compare metrics is according to the way they \emph{rank} different examples, rather than the exact scores.

\section{Additional Examples}
\label{sec:additional_examples}
In this section, we provide additional examples in which \cbs prefers the functionally correct prediction, while the best baseline metric in each language ranks higher a functionally incorrect prediction, which is inequivalent to the reference. 
\autoref{fig:java_example} shows an example in Java, and \autoref{fig:cpp_example} shows a C++ example.
\begin{figure*}[t]
    \newcommand{\nlsize}{0.36\textwidth}
    \newcommand{\lsize}{0.61\textwidth}
    \newcommand{\msize}{0.44\textwidth}
    \newcommand{\rsize}{0.54\textwidth}  
    \centering
    \begin{minipage}{\nlsize}
        \centering
        \textbf{Natural Language Question:} 
    \end{minipage}
    \hfill
  \begin{minipage}{\lsize}
  \centering
  \textbf{Reference:} 
  \end{minipage}\\
\begin{minipage}{\nlsize}
    \centering
    \begin{minted}    [fontsize=\footnotesize,stripnl=false,framesep=1pt,frame=single,breaksymbolleft=\;,escapeinside=||]{java}
/**
Find how many times a given 
substring can be found in 
the original string. 
Count overlaping cases.
>>> howManyTimes("", "a")
0
>>> howManyTimes("aaa", "a")
3
>>> howManyTimes("aaaa", "aa")
3
*/
\end{minted}
\end{minipage}
\hfill
\begin{minipage}{\lsize}
    \centering
  \begin{minted}[fontsize=\footnotesize,stripnl=false,framesep=1pt,frame=single,breaksymbolleft=\;,escapeinside=||]{java}
public static int howManyTimes(String string, 
      String substring) {
  int times = 0;
  for (int i = 0; i < string.length() - 
        substring.length() + 1; i++) {
    if (string.substring(i, i + substring.length())
        .equals(substring)) {
      times += 1;
    }
  }
  return times;
}
  \end{minted}
  \end{minipage} \\
  \begin{minipage}{\nlsize}
    \centering
    \begin{subfigure}[b]{1\linewidth}
    \caption{The natural language question.}
    \label{subfig:javanl}
    \end{subfigure}
    \end{minipage} 
  \begin{minipage}{\lsize}
    \centering
    \begin{subfigure}[b]{1\linewidth}
    \caption{The ground truth reference.}
    \label{subfig:javareference}
    \end{subfigure}
    \end{minipage} \\
    \vspace{2mm}
  \begin{minipage}{\msize}
    \centering
    \textbf{Non-equivalent candidate:} 
    \end{minipage} 
    \hfill
    \begin{minipage}{\rsize}
    \centering
    \textbf{Equivalent candidate:} 
    \end{minipage} \\
  \begin{minipage}{\msize}
  \begin{minted}[fontsize=\footnotesize,stripnl=false,framesep=1pt,frame=single,breaksymbolleft=\;,escapeinside=||]{java}
public static int howManyTimes(
  String string, String substring) {
  int count = 0;
  int index = 0;
  while ((index = string.indexOf(
        substring, index)) != -1) {
    count++;
    index += substring.length();
  }
  return count;
}

|\phantom{abc}|
  \end{minted}
  \end{minipage}
  \hfill
  \begin{minipage}{\rsize}
  \begin{minted}[fontsize=\footnotesize,framesep=1pt,stripnl=false,frame=single,breaksymbolleft=\;,escapeinside=||]{java}
public static int howManyTimes(
    String string, String substring) {
  int counter = 0;
  int index = 0;
  while (true) {
    index = string.indexOf(substring, index);
    if (index == -1)
      break;
    counter += 1;
    index += 1;
  }
  return counter;
}
  \end{minted}
  \end{minipage}
  
  \begin{minipage}{\msize}
  \centering
  \begin{subfigure}[b]{1\linewidth}
  \caption{\textbf{\textcolor{red}{Preferred by METEOR}}.}
  \label{subfig:javanoneq}
  \end{subfigure}
  \end{minipage}
  \hfill
  \begin{minipage}{\rsize}
  \centering
  \begin{subfigure}[b]{1\linewidth}
  \centering
  \caption{\textbf{\textcolor{ForestGreen}{Preferred by \cbs}}.}
  \label{subfig:javaeq}
  \end{subfigure}
  \end{minipage}
  \caption{An example in \he-Java, in which METEOR assigns a higher score to \Cref{subfig:javanoneq} which is \emph{not} functionally equivalent to the reference (\Cref{subfig:javareference}), while \cbs prefers the functionally correct candidate \Cref{subfig:javaeq}.
  }
  \label{fig:java_example}
  \end{figure*}

\begin{figure*}[t]
    \newcommand{\nlsize}{0.48\textwidth}
    \newcommand{\nlspace}{\hspace{0.05\textwidth}}
    \newcommand{\lsize}{0.45\textwidth}
    \newcommand{\msize}{0.45\textwidth}
    \newcommand{\rsize}{0.45\textwidth}  
    \centering
    \begin{minipage}{\nlsize}
        \centering
        \textbf{Natural Language Question:} 
    \end{minipage}
    \nlspace
  \begin{minipage}{\lsize}
  \centering
  \textbf{Reference:} 
  \end{minipage}\\
  \begin{minipage}{\nlsize}
    \centering
    \begin{minted}    [fontsize=\footnotesize,stripnl=false,framesep=1pt,frame=single,breaksymbolleft=\;,escapeinside=||]{cpp}
/*
Write a function that takes a vector 
of numbers as input and returns the
number of elements in the vector that
are greater than 10 and both first 
and last digits of a number are odd 
(1, 3, 5, 7, 9). For example:
specialFilter({15, -73, 14, -15}) 
    => 1 
specialFilter({33, -2, -3, 45, 21, 109}) 
    => 2
*/
    \end{minted}
\end{minipage}
\nlspace
\begin{minipage}{\lsize}
    \centering
  \begin{minted}[fontsize=\footnotesize,stripnl=false,framesep=1pt,frame=single,breaksymbolleft=\;,escapeinside=||]{cpp}
long specialFilter(
    std::vector<long> nums) {
  int num=0;
  for (int i=0;i<nums.size();i++)
    if (nums[i]>10) {
        string w=to_string(nums[i]);
        if (w[0]%2==1 and 
            w[w.length()-1] % 2 == 1) 
          num+=1;
    }
  return num;
}
  \end{minted}
  \end{minipage} \\
  \begin{minipage}{\nlsize}
    \centering
    \begin{subfigure}[b]{1\linewidth}
    \caption{The natural language question.}
    \label{subfig:cppnl}
    \end{subfigure}
    \end{minipage} 
  \begin{minipage}{\lsize}
    \centering
    \begin{subfigure}[b]{1\linewidth}
    \caption{The ground truth reference.}
    \label{subfig:cppreference}
    \end{subfigure}
    \end{minipage} \\
    \vspace{2mm}
  \begin{minipage}{\msize}
    \centering
    \textbf{Non-equivalent candidate:} 
    \end{minipage} 
    \nlspace
    \begin{minipage}{\rsize}
    \centering
    \textbf{Equivalent candidate:} 
    \end{minipage} \\
  \begin{minipage}{\msize}
  \begin{minted}[fontsize=\footnotesize,stripnl=false,framesep=1pt,frame=single,breaksymbolleft=\;]{cpp}
long specialFilter(
    std::vector<long> nums) {
  // Write your code here
  long sum = 0;
  for(int i=0; i<nums.size(); i++){
    std::string str = 
        std::to_string(nums[i]);
    long firstDigit = nums[i] / 
        abs(nums[i]);
    long lastDigit = nums[i]%10;
    if(nums[i] > 10 && 
        (firstDigit % 2 == 1 || 
            lastDigit % 2 == 1)){
      sum++;
    }
  }
  return sum;
}
\end{minted}
  \end{minipage}
  \nlspace
  \begin{minipage}{\rsize}
  \begin{minted}[fontsize=\footnotesize,framesep=1pt,stripnl=false,frame=single,breaksymbolleft=\;,escapeinside=||]{cpp}
long specialFilter(
    std::vector<long> nums) {
  long count = 0;
  for(int i=0; i<nums.size(); i++){
    if(nums[i] >= 10) {
      int last = nums[i] % 10;
      int first = nums[i];
      while(first >= 10) {
        first /= 10;
      }
      if (last % 2 != 0 && 
          first % 2 != 0) {
        count++;
      }
    }
  }
  return count;
}
  \end{minted}
  \end{minipage}
  
  \begin{minipage}{\msize}
  \centering
  \begin{subfigure}[b]{1\linewidth}
  \caption{\textbf{\textcolor{red}{Preferred by chrF}}.}
  \label{subfig:cppnoneq}
  \end{subfigure}
  \end{minipage}
  \nlspace
  \begin{minipage}{\rsize}
  \centering
  \begin{subfigure}[b]{1\linewidth}
  \centering
  \caption{\textbf{\textcolor{ForestGreen}{Preferred by \cbs}}.}
  \label{subfig:cppeq}
  \end{subfigure}
  \end{minipage}
  \caption{
    An example in \he-C++, in which chrF assigns a higher score to \Cref{subfig:cppnoneq} which is \emph{not} functionally equivalent to the reference (\Cref{subfig:cppreference}), while \cbs assigns a higher score to the functionally correct candidate \Cref{subfig:cppeq}.
  }
  \label{fig:cpp_example}
  \end{figure*}

\end{document}